# Crossover between bulk and interface photovoltaic mechanisms in ferroelectric vertical heterostructure

Amr Abdelsamie,[1] Lu You,[2,*] Le Wang,[3] Shuzhou Li,[1] Mingqiang Gu,[4] and Junling Wang[1,4,*]

[1]School of Materials Science and Engineering, Nanyang Technological University, Singapore 639798

[2]Jiangsu Key Laboratory of Thin Films, School of Physical Science and Technology, Soochow University, 1 Shizi Street, Suzhou 215006, China

[3]Physical and Computational Sciences Directorate, Pacific Northwest National Laboratory, Richland, WA, 99354, USA

[4]Department of Physics, Southern University of Science and Technology, Shenzhen 518055, China

Bulk photovoltaic (BPVE) effect in crystals lacking inversion symmetry offers great potential for optoelectronic applications due to its unique properties such as above bandgap photovoltage and switchable photocurrent. Because of their large spontaneous polarizations, ferroelectric materials are ideal platforms for studying BPVE. However, identifying the origin of experimentally observed photovoltaic response is often challenging due to the entanglement between bulk and interface effects, leading to much debate in the field. This issue is particularly pronounced in vertical heterostructures, where the two effects are comparable. Here we report a crossover between bulk- and interface-dominant response in vertical $BiFeO_3$ heterostructures when changing the photon energy. We show that well above-bandgap excitation leads to bulk photovoltaic response, but band-edge excitation requires interface band bending to separate the photocarriers. Our findings not only help to clarify contradicting reports in the literature, but also lay the ground for a deeper understanding of ferroelectric photovoltaic effect and its applications in various devices.

## I. INTRODUCTION

Bulk photovoltaic effect (BPVE), which occurs in noncentrosymmetric crystals such as ferroelectrics, is one of the most striking manifestations of non-linear optical phenomena [1]. Under uniform illumination, photocurrent (or photovoltage) is induced in otherwise spatially homogeneous crystals by taking advantage of the crystal asymmetry to separate the photoexcited electron-hole (e-h) pairs [Fig. 1(a)] [2-4]. The photocurrent is generated throughout the active layer and above bandgap photovoltage can be achieved, unlike in conventional junction-based photovoltaics [5-7]. Thus, solar cells based on BPVE are promising alternatives to overcome the Shockley-Queisser (SQ) efficiency limit [8]. However, it is generally accepted that Schottky barriers at the ferroelectric/metal interfaces can also lead to charge separation [Fig. 1(b) and 1(c)] [9-11]. In both cases, photovoltaic responses in ferroelectric-based cells are bidirectional following the polarization reversal [Fig. 1(d)] [12-15]. Furthermore, it has been argued that ferroelectric domain walls (DWs) may act as photo-electromotive sources to separate e-h pairs [5,16,17] or introduce additional conduction path [18,19].

One fingerprint of BPVE is the angular dependence of photovoltaic response on the light polarization [4,20]. This property, in turn, has been exploited to distinguish the BPVE from the conventional one. To date, most of the studies are carried out on ferroelectrics in either single crystal form [2,21,22] or thin films with co-planar electrodes [17,23-25]. In these geometries, the active layer size is very large, diminishing the impact of interface Schottky barrier on the total photovoltaic output. However, vertical heterostructures are much more relevant to applications. In this case, the interface depletion region would be comparable in dimension to the ferroelectric layer thickness. There are large volume of studies concluding that photovoltaic response in vertical heterostructures is dominated by the interface band bending [12,14,26].

However, there also exist a number of reports on BPVE in vertical heterostructures comprising BiFeO$_3$ [27], BaTiO$_3$ [28,29], and BiVO$_4$ [30], where contributions from interfaces or the tensorial nature of BPVE were not thoroughly examined. Therefore, clarifying the dominating photovoltaic mechanism in vertical ferroelectric heterostructures is of crucial importance for both fundamental study and practical applications.

Bismuth ferrite (BiFeO$_3$) offers an ideal platform to investigate the ferroelectric photovoltaic effect. It crystallizes in the rhombohedral space group R3c (pseudo-cubic (pc) lattice parameters $a_{pc}$ = 3.965 Å, $\alpha_{pc}$ = 89.4°) and possesses a spontaneous polarization along [111]$_{pc}$ (remnant polarization $P_r$ ~ 100 μC cm$^{-2}$) [31]. In this study, we aim to clarify the dominating mechanism of photovoltaic response in vertical BiFeO$_3$ capacitors by exploiting linearly polarized lights. It has been shown that BiFeO$_3$ exhibits uniaxial optical anisotropy which coincides with the polar axis, i.e., [111]$_{pc}$. In other words, linearly polarized light would undergo anisotropic absorption in BiFeO$_3$ [32-34], which will be taken into consideration. Furthermore, single-domain BiFeO$_3$ thin films are used to exclude the impact of domain walls.

## II. EXPERIMENTS
### A. Heterostructure preparation

Our device consists of a 500 nm (001)$_{pc}$-oriented BiFeO$_3$ film sandwiched between La$_{0.7}$Sr$_{0.3}$MnO$_3$ (10 nm, bottom electrode) and Pt (10 nm, top electrode). BiFeO$_3$ and La$_{0.7}$Sr$_{0.3}$MnO$_3$ were epitaxially grown using pulsed laser deposition (PLD) technique. In order to obtain a single domain state, the films were deposited on (001)-oriented SrTiO$_3$ substrates with 4° miscut towards <110>$_{pc}$ direction [35]. For the absorption measurements, two-side-polished SrTiO$_3$ substrates were used. Stoichiometric targets were ablated by a KrF excimer laser (248 nm). The growth parameters are summarized in Table (I). Following the deposition, arrays of 40×40 $\mu m^2$ Pt electrodes were sputtered on top of BiFeO$_3$ films through a shadow mask at room temperature.

TABLE I. Growth conditions for BiFeO$_3$ and La$_{0.7}$Sr$_{0.3}$MnO$_3$

| Thin film | Substrate temperature (°C) | Oxygen pressure (mTorr) | Repetition rate (Hz) | Fluence (J cm$^{-2}$) | Thickness (nm) |
|---|---|---|---|---|---|
| BiFeO$_3$ | 650 | 50 | 10 | 1.2 | 500 |
| La$_{0.7}$Sr$_{0.3}$MnO$_3$ | 800 | 200 | 3 | 2 | 10 |

**B. Device characterization**

A ferroelectric tester (Precision Multiferroic, Radiant Technologies) was used to measure the ferroelectric hysteresis loop. The local polarization mapping and switching were performed using a piezoelectric force microscope (Asylum Research MFP-3D) with Pt/Ir-coated tips. The I-V data were collected by using a pA metre/direct current (DC) voltage source (Hewlett Package 4140B) on a low noise probe station. To illuminate the cell, lasers with wavelengths of 405 nm (~ 0.65 W cm$^{-2}$) and 520 nm (~ 0.19 W cm$^{-2}$) were used. White light was provided by a halogen lamp with an intensity of about 0.05 W cm$^{-2}$. A Glan-Thomson calcite polarizer is placed between the objective lens and the light source to obtain linearly polarized light.

**C. Linear dichroism in BiFeO$_3$**

The optical absorption of BiFeO$_3$ films were determined using a Perkin Elmer Lambda 950 UV-vis spectrophotometer. BiFeO$_3$ films (170 nm) were directly grown on double-side-polished SrTiO$_3$ substrates to ensure transparency. All curves were collected for wavelengths between 400 nm and 600 nm at room temperature. It should be noted that accurate data for wavelengths shorter than 400 nm could not be obtained due to substrate absorption. The data were deduced from absorbance spectra after removing contributions from the SrTiO$_3$ substrate.

To study the absorption of BiFeO$_3$ films for linearly polarized light, the spectrometer was equipped with a Glan-Thomson calcite polarizer (resembling the one used to measure the photovoltaic performance). Moreover, the sample was mounted on a rotatable stand to enable rotation between -90º and +90º while ensuring a constant light intensity.

## III. RESULTS
### A. Characterizing ferroelectricity and optical property

Prior to the photovoltaic measurements, we first assess the basic ferroelectric and optical properties of the BiFeO$_3$ active layer. BiFeO$_3$ exhibits a single ferroelectric domain structure, which is confirmed by piezoelectric force microscope (PFM) as shown in Fig. 1(e). The BiFeO$_3$ is single phase with a c-axis lattice constant of 3.978 Å. Figure 1(f) presents the ferroelectric hysteresis loop and transient current curve. The rectangular shape of the loop reflects the intrinsic and monodomain ferroelectricity of the BiFeO$_3$ film with a remnant polarization of P$_r$ = 65 μC cm$^{-2}$ along [001]$_{pc}$. The optical absorption spectrum of the BiFeO$_3$ film [Fig. 1(g)] shows a direct gap at about 2.7 eV (~460 nm) as plotted in Fig. S1 in Supplemental Material, which is attributed to $t_{1u}(\pi) \rightarrow t_{2g}(\pi^*)$ dipole-allowed O-2p to Fe-3d charge transfer (CT) transition [32,36]. It also displays a weak yet distinct absorption peak with the onset at about 2.2 eV (~560 nm), which is usually assigned to the $t_{1g}(\pi) \rightarrow t_{2g}(\pi^*)$ dipole-forbidden CT transition. This low-lying electronic structure has been widely reported in BiFeO$_3$ samples in various forms and by different synthesis methods, and it appears to be a common feature of CT ferrite insulators with low-symmetry FeO$_6$ octahedra distortions, signifying the intrinsic nature of this transition [36]. Furthermore, due to the strong electron-phonon coupling, the sub-bandgap transition results in local lattice deformation and thus the formation of self-trapped excitonic states [37,38], which also underlies the previously reported ultrafast photostriction [39,40] and broadband photoluminescence emission [41] of BiFeO$_3$.

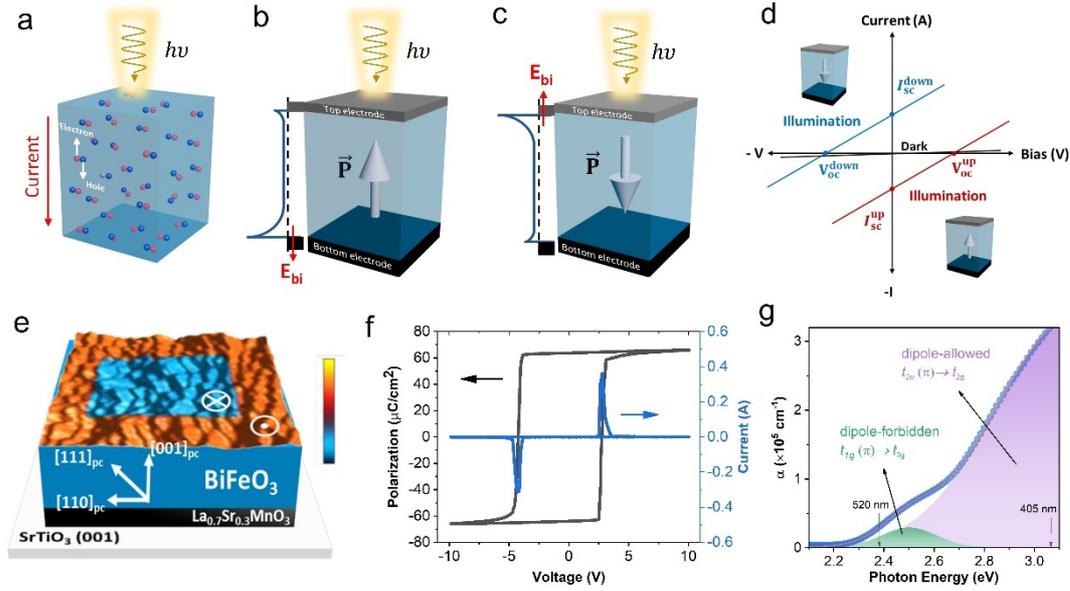

Fig. 1. Photovoltaic effects in ferroelectric materials and characteristics of the BiFeO$_3$ films used in this study. a) Photogenerated electron-hole pairs in noncentrosymmetric crystals are separated in the homogeneous bulk yielding bulk photovoltaic effect. However, bringing metallic electrodes (grey and black layers) in contact with the ferroelectric material creates Schottky barriers. Switching the ferroelectric polarization $\vec{P}$ from b) up-state to c) down-state modulates the Schottky barrier heights at the top and bottom interfaces. The electronic band structure is depicted on the left-hand side of the device (dashed line represents Fermi level). The photoexcited charge carries are separated at the interface. d) Reversal of the ferroelectric polarization flips the photocurrent between negative and positive directions. e) Piezoresponse force microscopy phase image overlaid on the topography of BiFeO$_3$ film with as-grown and switched areas. f) Polarization-voltage and switching current curves of the Pt/BiFeO$_3$/La$_{0.7}$Sr$_{0.3}$MnO$_3$ capacitor. g) Measured absorption coefficient of BiFeO$_3$, revealing the weak dipole-forbidden charge transfer (CT) transition with the onset at 2.2 eV and the strong dipole-allowed CT transition above the nominal bandgap of 2.7 eV. Purple and green arrows point to the laser energies used in this study.

**B. Light polarization angle dependent photocurrents**

Photovoltaic measurements were performed on Pt/BiFeO$_3$/La$_{0.7}$Sr$_{0.3}$MnO$_3$ devices using linearly polarized lights. Here, 405-nm (~ 3.1 eV) and 520-nm (~ 2.4 eV) lasers with maximum intensities of 0.65 W cm$^{-2}$ and 0.19 W cm$^{-2}$, and a halogen lamp (maximum intensity of 50 mW cm$^{-2}$) were used as the light sources, respectively. The sample was first poled into polarization up or down state. A linear Glan-Thomson calcite polarizer was placed between the sample and the focusing lens and the sample was illuminated through the top electrode (the thin Pt electrode allows 37%, 32% and 23% of the 405 nm, 520 nm and white lights, respectively, to pass through). As plotted in Fig. 2(a), the light polarization makes an angle $\theta$ relative to the in-plane ferroelectric polarization (P$_{in}$) of BiFeO$_3$, i.e. [110]$_{pc}$. Hence, $\theta = 0°$ is at which light polarization is *parallel* to the in-plane ferroelectric polarization. The out-of-plane photovoltaic behavior was measured for different azimuthal angles of the light polarization (from -90° to +90°). The short-circuited currents in both polarization-up and -down states are shown in Fig. 2(b) and 2(c) (more data and discussion can be found in Supplemental Material).

Interestingly, modulation of the photocurrents with respect to $\theta$ is observed for both 405 nm (in purple) and 520 nm (in green) lights, but in completely opposite manners. Under 405 nm illumination, the photocurrent exhibits sinusoidal behavior with its maximum at $\theta = 0°$ and minimum at $\theta = \pm 90°$, which even flips its direction for the up state. In contrast, when illuminating the device with 520 nm laser, the photocurrent direction remains negative for the up state and positive for the down state across all the angles, and contrarily, the sinusoidal photocurrent is maximized at $\theta = \pm 90°$. Unlike the 405 nm excitation, the 520 nm case shows almost symmetrical photocurrent response for opposite polarization states. The results suggest that, (i) there is a strong correlation between the photovoltaic response and the angle $\theta$; and (ii) different photovoltaic mechanisms may be at work for different photon energies.

It has been established that BPVE is described by a tensor of third order [4]. In particular, when illuminating (001)$_{pc}$-oriented BiFeO$_3$ crystal, the bulk photocurrent is given by (details in Methods):

$$J^{h\nu}_{[001]}(\theta) = \frac{I_{opt}}{3\sqrt{3}}[A + B\sin(2\theta + \frac{\pi}{2})] \qquad (1)$$

where $J^{h\nu}_{Bulk}$ is the photocurrent density along [001]$_{pc}$ under $h\nu$ illumination, $I_{opt}$ is the light intensity, $A$ and $B$ are functions of photovoltaic tensor $\beta_{ij}$ of BiFeO$_3$ and $\theta$ is the angle between light polarization and $P_{in}$ of BiFeO$_3$. We then calculated $I^{405nm}_{[001]}$ and $I^{520nm}_{[001]}$, in which $\beta_{ij}$ values were extracted from experimental work on BiFeO$_3$ thin films with co-planar configurations, i.e., from measured bulk photocurrents as listed in Table (AI) [taken from ref. (25)]. Apparently, the calculated bulk photocurrent under 405 nm light [Fig. 2(d)] reveals a trend (sinusoidal shape) and magnitude coinciding with the experimental result shown in Fig. 2(b), though a large vertical shift is evident for the polarization down state. In contrast, the photocurrent under 520 nm laser reveals a trend totally opposite to that described by Eq. (1), and the $I_{sc}$ value is orders of magnitude larger than that calculated for BPVE. However, it can nonetheless be fitted by a sinusoidal function. Furthermore, linearly polarized white light (generated by a halogen lamp) yields a similar behavior as that of 520 nm laser (Supplemental Material).

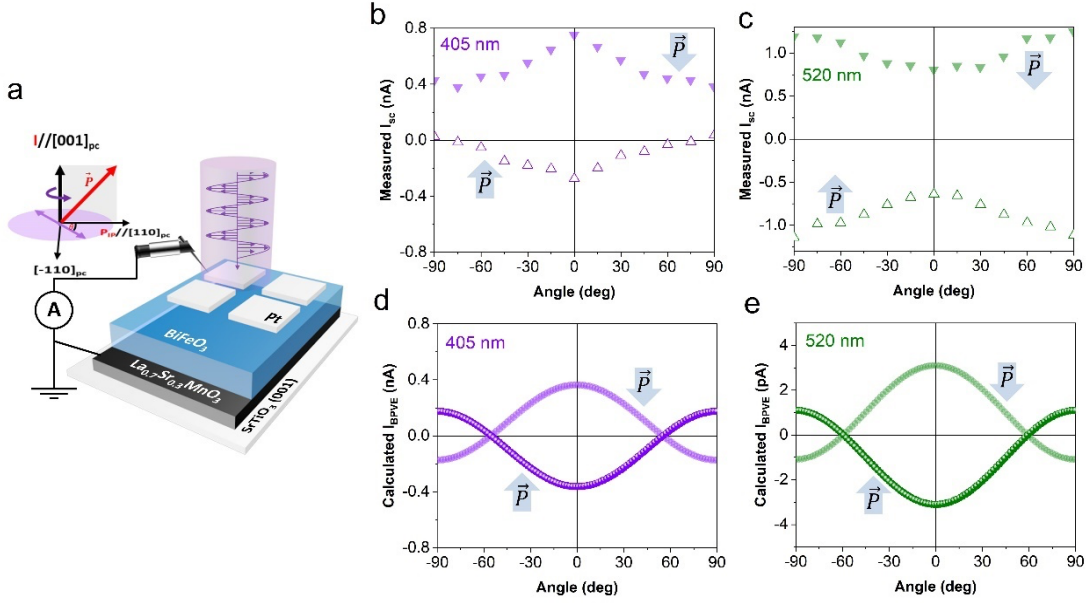

Fig. 2. Light polarization-dependent photovoltaic response of vertical Pt/BiFeO$_3$/La$_{0.7}$Sr$_{0.3}$MnO$_3$ capacitors. a) Schematic diagram of the experimental setup. The light polarization angle $\theta$ is defined with respect to the in-plane ferroelectric polarization of the sample (i.e., [110]$_{pc}$). Modulation of photocurrents in polarization-up and -down states collected under b) 405 nm (~3.1 eV) and c) 520 nm (~2.4 eV) excitations. Calculated I$_{BPVE}$ under d) 405 nm (~3.1 eV) and e) 520 nm (~2.4 eV) excitations using Eq. (1).

The apparent opposite behavior suggests that the governing photovoltaic mechanisms for 405 nm and 520 nm lights are likely different. While the 405 nm light likely generates BPVE, what is causing the sinusoidal photocurrent under 520 nm illumination? One possibility is the anisotropic light absorption of BiFeO$_3$ [42]. We thus measured the light polarization dependent absorption of monodomain BiFeO$_3$ films using a UV-vis spectrophotometer to evaluate the absorption anisotropy quantitatively.

## C. Optical linear dichroism of BiFeO$_3$ films

Figure 3 displays the light polarization dependent absorption of a BiFeO$_3$ film for 0º < $\theta$ < 90º after excluding contributions from the substrate. As shown in Fig. 3(c), the percentage of variation between the minimum and maximum absorption, given by $[A(90º) - A(0º)/A(90º)]$%, exhibits positive value over the whole energy range. This indicates that the absorption reaches maximum when the light and ferroelectric polarizations are orthogonal, and minimum when parallel. In order to correlate the anisotropic absorption with the photovoltaic response, the angle-dependent absorption at 405 nm and 520 nm were extracted [Fig. 3(d) and 3(e)], which reveal angular modulation of around 33% and 36%, respectively. Furthermore, the angle-dependent integrated area under the absorption curves is also plotted in Fig. 3(f), representing white light absorption anisotropy in BiFeO$_3$ (percentage of modulation ~29%). The light polarization dependence can be fitted by the following equation:

$$Absorption = a + b \cdot sin(2\theta - \frac{\pi}{2} + \varphi) \qquad (2)$$

where *a* and *b* are positive constants and $\varphi$ accounts for experimental error in the rotation angle.

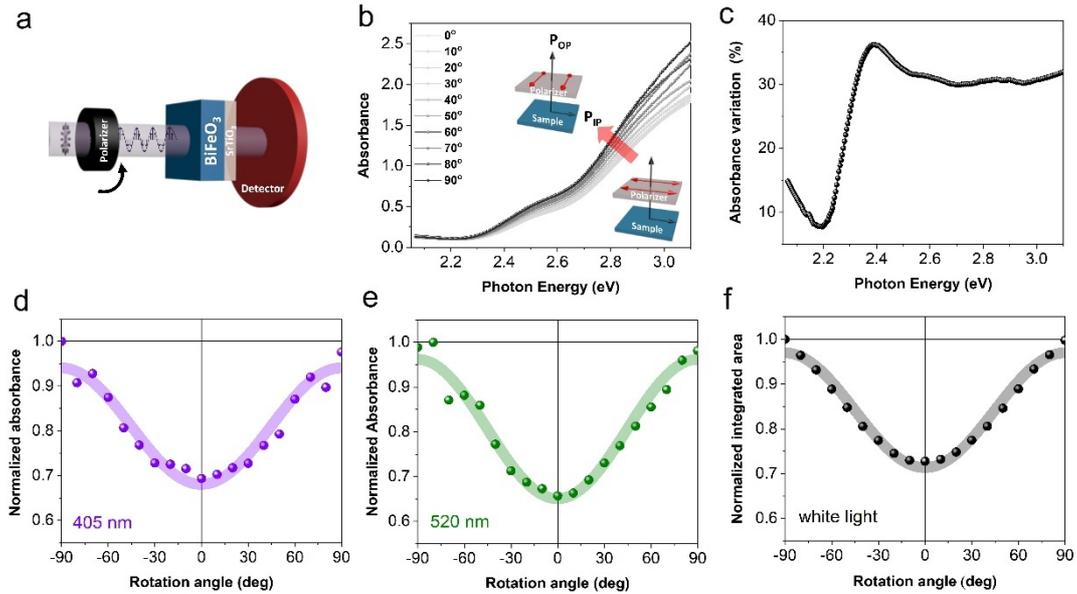

Fig. 3. Angle-dependent polarized light absorption spectra of BiFeO$_3$ on SrTiO$_3$. a) Schematic of the measurement set-up. b) Polarized light absorption spectra collected from 0º to +90º. c) Photon energy dependent absorbance variation, $[A(90º) - A(0º)/A(90º)]\%$ between two orthogonal light polarizations. d) Extracted angle-dependent absorption at 405 nm and 520 nm, and the total integrated area under the absorption curves representing the linear dichroism for white light.

It is shown that, when the light polarization is set along [110]$_{pc}$ of the (001)$_{pc}$-oriented BiFeO$_3$, the optical absorption is at its minimum. The optical excitation increases as the rotation angle increases and the strongest absorption occurs at light polarization perpendicular to [110]$_{pc}$. The absorption anisotropy is consistent with the band structure of BiFeO$_3$ (R3c) [43,44] and is attributed to the highly-distorted FeO$_6$ octahedra.

Coming back to the photovoltaic response, it is now necessary to take the anisotropic absorption into consideration. In order to handle this, the modulated photocurrent is normalized (divided) by the anisotropic absorption. In Fig. 4(f), the photocurrents under 520 nm laser become almost constant with respect to the light polarization angle once normalized by the light absorption anisotropy. Similar behavior is also found for white light illumination (see Fig. S3 in Supplemental Material). It suggests that in these two cases, the photocurrents originate mainly from interface effect, instead of BPVE. As for 405 nm excitation, the experimental data is only well fitted if an absorption-modulated interfacial term, I$_{Interface}$, is added to the BPVE current calculated based on Eq. (1). Hence, the total photocurrent under 405 nm light is given by:

$$I_{total}^{405nm}(\theta) = \pm I_{Interface} \pm 0.15 I_{Interface} \sin(2\theta - \frac{\pi}{2}) \pm 10^{-10}[0.93 + 2.66 \sin\left(2\theta + \frac{\pi}{2}\right)] \quad (3)$$

where the second term accounts for the modulation due to absorption anisotropy and the negative and positive signs corresponds to polarization up and down states, respectively.

The fitted results are shown in Fig. 4(c), where we obtain the $I_{Interface}$ values of -50 pA and +0.44 nA for up and down states, respectively. That means the contribution from interface effect is negligible in the polarization-up state, but becomes considerable in the down state, as indicated by the larger vertical shift of the photocurrent curve to the positive direction. The different interface contribution can be understood by the small absorption depth (~ 35 nm) at this wavelength (see Fig. S4 in Supplemental Material). In the polarization-down state, the top interface (Pt/BiFeO$_3$) is activated and directly exposed to the light which allows strong interface effect [Fig. 4(b)]. In the polarization-up state, the bottom interface (BiFeO$_3$/La$_{0.7}$Sr$_{0.3}$MnO$_3$) is activated, but with little light travelling through the whole film thickness [45], greatly reducing its contribution to the total photocurrent [Fig. 4(a)].

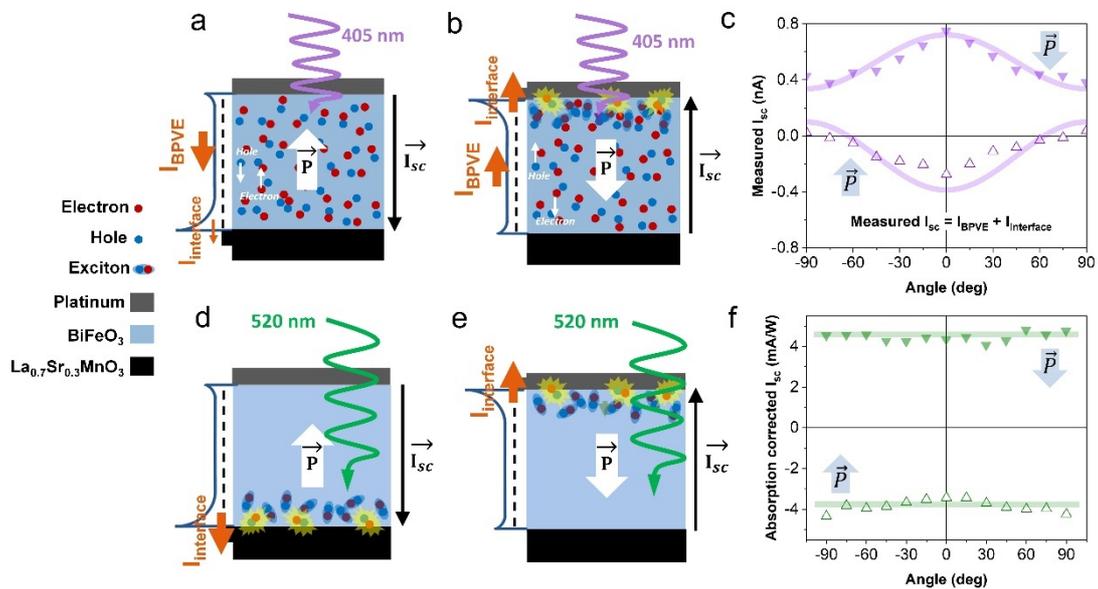

Fig. 4. The crossover between bulk and interface photovoltaic mechanisms. a) 405 nm laser preferentially generates bulk photovoltaic effects in polarization-up state with negligible current contribution from interface effect. b) In down-state, the interface contribution becomes

considerable since the top interface is activated and directly exposed to the illuminating light. c) Angle-dependent measured $I_{sc}$ (triangular dots) under 405 nm light fitted by considering combined bulk and interface effects (solid lines). Illumination by 520 nm laser in d) polarization-up and e, polarization-down state generates self-trapped excitons in the bulk of $BiFeO_3$, those around the interface are separated by interfacial built-in field and produce the photocurrent. f) Angle-dependent $I_{sc}$ (triangular dots) under 520 nm light are almost constant after normalized by the anisotropic absorption.

## IV. DISCUSSION

The model we proposed is consistent with our current understanding on the microscopic origins of BPVE in noncentrosymmetric materials, namely ballistic and shift currents. Ballistic photovoltaic current originates from the asymmetric distribution of hot photocarrier momenta in the reciprocal space [46]. By definition, high energy excitation (405 nm excitation in our case) is needed to provide excessive kinetic energy to the photocarriers. Shift current is associated with the shift of electron wave functions in asymmetric media under persistent illumination [47]. First-principles calculations have shown that transitions involving localized *d* orbitals exhibit small shift currents [48]. Since the bottom edge of $BiFeO_3$ conduction band is dominated by the Fe-3d states [49,50], band-edge excitation generates self-trapped excitonic states, unfavorable for the shift current. Hence, in both mechanisms, 520 nm excitation is unfavorable to produce large BPVE current. Nevertheless, the polarization-modulated band-bending field at the interface is efficient in dissociating the self-trapped excitons and results in a sizable photocurrent, which also shows the angular modulation due to absorption anisotropy.

## V. CONCLUSION

In summary, the dominating photovoltaic mechanism in monodomain $BiFeO_3$ based vertical heterostructures depends on the nature of the optical excitation. Under high energy excitation, the e-h pair separation is dominated by the crystal asymmetry of $BiFeO_3$, i.e., BPVE, in which the interface band-bending-driven photovoltaic effect is still present (which interface matters more depends on the ferroelectric polarization direction as shown in Fig. (4). On the other hand, the optical structure of $BiFeO_3$ is also characterized by self-trapped *p-d* charge transfer excitons near the absorption edge. These strongly coupled excitons require a strong external field, i.e., band-bending at the $BiFeO_3$/metal interface in our devices, to dissociate. Therefore, band-edge excitation leads to interface-driven photovoltaic response. In this case, the light polarization-dependent absorption also modulates the photocurrents. Both bulk and interface-driven photovoltaic effects give rise to sinusoidal dependence on the angle between light polarization and ferroelectric polarization of the sample, but with 90º phase difference, which allows us to distinguish the two effects. These findings help to disentangle the bulk and interface effects and significantly advance our understanding of ferroelectric photovoltaic effect in practical devices.

## ACKNOWLEDGEMENTS

L.Y. acknowledges the startup funds from Soochow University, and the support from Priority Academic Program Development (PAPD) of Jiangsu Higher Education Institutions. L.Y. also acknowledges the support from the National Natural Science Foundation of China (11774249, 12074278), the Natural Science Foundation of Jiangsu Province (BK20171209), and the Key University Science Research Project of Jiangsu Province (18KJA140004, 20KJA140001). J.W. acknowledges the support from the Ministry of Education, Singapore (Grant numbers: AcRF Tier 1 189/18), the startup grant from Southern University of Science and Technology (SUSTech), China and support from the National Natural Science Foundation of China (12074164).

# APPENDIX: CALCULATION OF THE PHOTOCURRENT INDUCED BY THE BULK PHOTOVOLTAIC EFFECT

When linearly polarized light is incident on a noncentrosymmetric crystal, the BPVE generated photocurrent is given by:

$$J_i = I_{opt}\beta_{ijk}e_j e_k \qquad (A1)$$

where $I_{opt}$ is the light intensity (see Supplemental Material for details), $\beta_{ijk}$ is a third-rank bulk photovoltaic tensor and $e_j$ and $e_k$ are the projections of light polarization vector. In this work, the light propagates along $z$-axis to the surface of a monodomain $(001)_{pc}$-BiFeO$_3$ thin film (space group of $R3c$) giving rise to a photocurrent:

$$J_i = I_o \begin{pmatrix} 0 & 0 & 0 & 0 & \beta_{15} & -\beta_{22} \\ -\beta_{22} & \beta_{22} & 0 & \beta_{15} & 0 & 0 \\ \beta_{31} & \beta_{31} & \beta_{33} & 0 & 0 & 0 \end{pmatrix} \begin{pmatrix} e_1^2 \\ e_2^2 \\ 0 \\ 0 \\ 0 \\ 2e_1 e_2 \end{pmatrix} \qquad (A2)$$

To be consistent with experimental measurements, the in-plane ferroelectric polarization is taken as reference. Therefore, the out-of-plane photocurrent can be written as:

$$J^{hv}_{[001]}(\theta) = \frac{I_{opt}}{3\sqrt{3}}[-2\beta_{15} - \sqrt{2}\beta_{22} + 2\beta_{31} + \beta_{33}] +$$

$$\frac{I_{opt}}{3\sqrt{3}}[-2\beta_{15} + 2\sqrt{2}\beta_{22} - \beta_{31} + \beta_{33}].\sin\left(2\theta + \frac{\pi}{2}\right), \qquad (A3)$$

Similarly, BPVE induced photocurrents along $[110]_{pc}$ and $[-110]_{pc}$ directions can be written as:

$$J^{hv}_{[110]}(\theta) = \frac{I_{opt}}{3\sqrt{3}}[\sqrt{2}\beta_{15} + \beta_{22} + 2\sqrt{2}\beta_{31} + \sqrt{2}\beta_{33}] +$$

$$\frac{I_{opt}}{3\sqrt{3}}[\sqrt{2}\beta_{15} - 2\beta_{22} - \sqrt{2}\beta_{31} + \sqrt{2}\beta_{33}].\sin\left(2\theta + \frac{\pi}{2}\right), \qquad (A4)$$

$$J^{hv}_{[-110]}(\theta) = I_{opt}\left[\frac{2}{\sqrt{6}}\beta_{15} + \frac{1}{\sqrt{3}}\beta_{22}\right].sin(2\theta). \tag{A5}$$

Where $\theta$ is the angle between the light polarization and in-plane ferroelectric polarization, i.e. $[110]_{pc}$. The $J^{hv}_{[001]}(\theta)$ shows a $sin(2\theta + \frac{\pi}{2})$ dependence. Moreover, $\beta_{ijk}$ values are wavelength dependent.

TABLE A1. Bulk photovoltaic tensor elements of BiFeO$_3$ at 405 nm (3.1 eV) and 520 nm (2.4 eV). Taken from Ref. (25).

| Light source | $\beta_{15}$ (V$^{-1}$) | $\beta_{22}$ (V$^{-1}$) | $\beta_{31}$ (V$^{-1}$) | $\beta_{33}$ (V$^{-1}$) |
|---|---|---|---|---|
| 405 nm (3.1 eV) | $8.1 \times 10^{-5}$ | $-1.1 \times 10^{-5}$ | $6.4 \times 10^{-5}$ | $-1.1 \times 10^{-4}$ |
| 520 nm (2.4 eV) | $3.0 \times 10^{-6}$ | $-4.4 \times 10^{-7}$ | $1.6 \times 10^{-6}$ | $-4.4 \times 10^{-6}$ |

Experimentally, photocurrents in $(001)_{pc}$-BiFeO$_3$ can be measured along the in-plane polarization direction ($[110]_{pc}$) and its perpendicular direction ($[-110]_{pc}$) by employing planar configuration. Matsuo *et al* performed such measurements and obtained the $\beta_{15}, \beta_{22}, \beta_{31}$ and $\beta_{33}$ values for 405 nm and 520 nm lasers [25].

In vertical configuration and by incorporating the experimental values of $\beta_{15}, \beta_{22}, \beta_{31}$ and $\beta_{33}$ (derived from planar configuration experiment) into Eq. (A3), the out-of-plane BPVE photocurrent under 405 nm (3.1 eV) and 520 nm (2.4 eV) lights are as follows,

$$I^{405nm}_{[001]}(\theta) = -I_{opt}.10^{-10}[3.952 + 11.296\,sin\left(2\theta + \frac{\pi}{2}\right)], \tag{A6}$$

and

$$I^{520nm}_{[001]}(\theta) = -I_{opt}.10^{-11}[2.025 + 4.06\,sin\left(2\theta + \frac{\pi}{2}\right)]. \tag{A7}$$

In polarization-up state, the $I_{sc}$ is negative at $\theta = 0°$, and flips polarity when the light polarization is perpendicular to the in-plane ferroelectric polarization. Both 405 nm- and 520 nm-light induced BPVE photocurrents follow the same trend, but $I_{[001]}^{405nm}$ is one order-of-magnitude larger than $I_{[001]}^{520nm}$ for the same light intensity. Furthermore, switching the ferroelectric polarization to down state changes the sign of $β_{ij}$ and the polarity of the photocurrent.

--------------------------------

*Corresponding author: jwang@sustech.edu.cn (J.W.); lyou@suda.edu.cn (L.Y.)

[1] B. I. Sturman, V. M. Fridkin, and J. Bradley, *The photovoltaic and photorefractive effects in noncentrosymmetric materials* (Routledge, 2021).

[2] T. Choi, S. Lee, Y. J. Choi, V. Kiryukhin, and S.-W. Cheong, Switchable Ferroelectric Diode and Photovoltaic Effect in BiFeO$_3$, Science **324**, 63 (2009).

[3] V. M. Fridkin, *Photoferroelectrics* (Springer Science & Business Media, 2012), Vol. 9.

[4] M.-M. Yang, D. J. Kim, and M. Alexe, Flexo-photovoltaic effect, Science **360**, 904 (2018).

[5] S. Yang, J. Seidel, S. Byrnes, P. Shafer, C.-H. Yang, M. Rossell, P. Yu, Y.-H. Chu, J. Scott, J. Ager, Above-bandgap voltages from ferroelectric photovoltaic devices, Nat. Nanotech. **5**, 143 (2010).

[6] M. Alexe and D. Hesse, Tip-enhanced photovoltaic effects in bismuth ferrite, Nat. Commun. **2**, 1 (2011).

[7] A. Bhatnagar, A. R. Chaudhuri, Y. H. Kim, D. Hesse, and M. Alexe, Role of domain walls in the abnormal photovoltaic effect in BiFeO$_3$, Nat. Commun. **4**, 1 (2013).

[8] W. Shockley and H. J. Queisser, Detailed balance limit of efficiency of p-n junction solar cells, J. Appl. Phys. **32**, 510 (1961).

[9] P. Blom, R. Wolf, J. Cillessen, and M. Krijn, Ferroelectric schottky diode, Phys. Rev. Lett. **73**, 2107 (1994).

[10] L. Pintilie and M. Alexe, Metal-ferroelectric-metal heterostructures with Schottky contacts. I. Influence of the ferroelectric properties, J. Appl. Phys. **98**, 124103 (2005).

[11] L. Pintilie, I. Boerasu, M. Gomes, T. Zhao, R. Ramesh, and M. Alexe, Metal-ferroelectric-metal structures with Schottky contacts. II. Analysis of the experimental current-voltage and capacitance-voltage characteristics of Pb (Zr, Ti)$O_3$ thin films, J. Appl. Phys. **98**, 124104 (2005).

[12] Z. Tan, L. Hong, Z. Fan, J. Tian, L. Zhang, Y. Jiang, Z. Hou, D. Chen, M. Qin, M. Zeng, Thinning ferroelectric films for high-efficiency photovoltaics based on the Schottky barrier effect, NPG Asia Mater. **11**, 1 (2019).

[13] R. Guo, L. You, Y. Zhou, Z. S. Lim, X. Zou, L. Chen, R. Ramesh, and J. Wang, Non-volatile memory based on the ferroelectric photovoltaic effect, Nat. Commun. **4**, 1 (2013).

[14] D. Lee, S. H. Baek, T. H. Kim, J. G. Yoon, C. M. Folkman, C. B. Eom, and T. W. Noh, Polarity control of carrier injection at ferroelectric/metal interfaces for electrically switchable diode and photovoltaic effects, Phys. Rev. B **84**, 125305 (2011).

[15] D. S. Knoche, Y. Yun, N. Ramakrishnegowda, L. Mühlenbein, X. Li, and A. Bhatnagar, Domain and switching control of the bulk photovoltaic effect in epitaxial BiFe$O_3$ thin films, Sci. Rep. **9**, 1 (2019).

[16] J. Seidel, D. Fu, S.-Y. Yang, E. Alarcón-Lladó, J. Wu, R. Ramesh, and J. W. Ager, Photovoltaic Current Generation at Ferroelectric Domain Walls, Phys. Rev. Lett. **107**, 126805 (2011).

[17] H. Matsuo, Y. Kitanaka, R. Inoue, Y. Noguchi, M. Miyayama, T. Kiguchi, and T. J. Konno, Bulk and domain-wall effects in ferroelectric photovoltaics, Phys. Rev. B **94**, 214111 (2016).

[18] S. Farokhipoor and B. Noheda, Conduction through 71° Domain Walls in BiFe$O_3$ Thin Films, Phys. Rev. Lett. **107**, 127601 (2011).

[19] Y. Zhou, L. Fang, L. You, P. Ren, L. Wang, and J. Wang, Photovoltaic property of domain engineered epitaxial BiFeO$_3$ films, Appl. Phys. Lett. **105**, 252903 (2014).

[20] V. M. Fridkin, Bulk photovoltaic effect in noncentrosymmetric crystals, Crys. Rep. **46**, 654 (2001).

[21] H. Festl, P. Hertel, E. Krätzig, and R. v. Baltz, Investigations of the photovoltaic tensor in doped LiNbO$_3$, Phys. Status Solidi B **113**, 157 (1982).

[22] V. M. Fridkin, Parity nonconservation and bulk photovoltaic effect in a crystal without symmetry center, IEEE Trans. Ultrason. Ferr. **60**, 1551 (2013).

[23] W. Ji, K. Yao, and Y. C. Liang, Evidence of bulk photovoltaic effect and large tensor coefficient in ferroelectric BiFeO$_3$ thin films, Phys. Rev. B **84**, 094115 (2011).

[24] M.-M. Yang, Z.-D. Luo, D. J. Kim, and M. Alexe, Bulk photovoltaic effect in monodomain BiFeO3 thin films, Appl. Phys. Lett. **110**, 183902 (2017).

[25] H. Matsuo, Y. Noguchi, and M. Miyayama, Gap-state engineering of visible-light-active ferroelectrics for photovoltaic applications, Nat. Commun. **8**, 1 (2017).

[26] R. Nechache, C. Harnagea, S. Li, L. Cardenas, W. Huang, J. Chakrabartty, and F. Rosei, Bandgap tuning of multiferroic oxide solar cells, Nat. Photon. **9**, 61 (2015).

[27] W. Ji, K. Yao, and Y. C. Liang, Bulk photovoltaic effect at visible wavelength in epitaxial ferroelectric BiFeO$_3$ thin films, Adv. Mater. **22**, 1763 (2010).

[28] A. Zenkevich, Y. Matveyev, K. Maksimova, R. Gaynutdinov, A. Tolstikhina, and V. Fridkin, Giant bulk photovoltaic effect in thin ferroelectric BaTiO$_3$ films, Phys. Rev. B **90**, 161409 (2014).

[29] J.E. Spanier, V.M. Fridkin, A.M. Rappe, A.R. Akbashev, A. Polemi, Y. Qi, Z. Gu, S.M. Young, C.J. Hawley, D. Imbrenda, Power conversion efficiency exceeding the Shockley–Queisser limit in a ferroelectric insulator, Nat. Photon. **10**, 611 (2016).

[30] H. Mai, T. Lu, Q. Sun, R.G. Elliman, F. Kremer, K. Catchpole, Q. Li, Z. Yi, T.J. Frankcombe, Y. Liu, High performance bulk photovoltaics in narrow-bandgap centrosymmetric ultrathin films, Mater. Horiz. **7**, 898 (2020).

[31] G. Catalan and J. F. Scott, Physics and applications of bismuth ferrite, Adv. Mater. **21**, 2463 (2009).

[32] D. Schmidt, L. You, X. Chi, J. Wang, and A. Rusydi, Anisotropic optical properties of rhombohedral and tetragonal thin film $BiFeO_3$ phases, Phys. Rev. B **92**, 075310 (2015).

[33] S. Choi, H. Yi, S.-W. Cheong, J. Hilfiker, R. France, and A. Norman, Optical anisotropy and charge-transfer transition energies in $BiFeO_3$ from 1.0 to 5.5 eV, Phys. Rev. B **83**, 100101 (2011).

[34] C. Tabares-muñoz, J.-P. Rivera, and H. Schmid, Ferroelectric domains, birefringence and absorption of single crystals of $BiFeO_3$, Ferroelectrics **55**, 235 (1984).

[35] L. You, F. Zheng, L. Fang, Y. Zhou, L.Z. Tan, Z. Zhang, G. Ma, D. Schmidt, A. Rusydi, L. Wang, Enhancing ferroelectric photovoltaic effect by polar order engineering, Sci. Adv. **4**, eaat3438 (2018).

[36] R. Pisarev, A. Moskvin, A. Kalashnikova, and T. Rasing, Charge transfer transitions in multiferroic $BiFeO_3$ and related ferrite insulators, Phys. Rev. B **79**, 235128 (2009).

[37] Y. Li, C. Adamo, C. E. Rowland, R. D. Schaller, D. G. Schlom, and D. A. Walko, Nanoscale excitonic photovoltaic mechanism in ferroelectric $BiFeO_3$ thin films, APL Mater. **6**, 084905 (2018).

[38] Y. Yamada, T. Nakamura, S. Yasui, H. Funakubo, and Y. Kanemitsu, Measurement of transient photoabsorption and photocurrent of $BiFeO_3$ thin films: Evidence for long-lived trapped photocarriers, Phys. Rev. B **89**, 035133 (2014).

[39] H. Wen, P. Chen, M.P. Cosgriff, D.A. Walko, J.H. Lee, C. Adamo, R.D. Schaller, J.F. Ihlefeld, E.M. Dufresne, D.G. Schlom, Electronic origin of ultrafast photoinduced strain in BiFeO$_3$, Phys. Rev. Lett. **110**, 037601 (2013).

[40] Y. Li, C. Adamo, P. Chen, P.G. Evans, S.M. Nakhmanson, W. Parker, C.E. Rowland, R.D. Schaller, D.G. Schlom, D.A. Walko, Giant optical enhancement of strain gradient in ferroelectric BiFeO$_3$ thin films and its physical origin, Sci. Rep. **5**, 1 (2015).

[41] Y.-M. Sheu, S. Trugman, Y.-S. Park, S. Lee, H. Yi, S.-W. Cheong, Q. Jia, A. Taylor, and R. Prasankumar, Ultrafast carrier dynamics and radiative recombination in multiferroic BiFeO$_3$, Appl. Phys. Lett. **100**, 242904 (2012).

[42] R. H. Bube, *Photoconductivity of solids* (RE Krieger Pub. Co., 1978).

[43] J. K. Shenton, D. R. Bowler, and W. L. Cheah, Influence of crystal structure on charge carrier effective masses in BiFeO$_3$, Phys. Rev. B **100**, 085120 (2019).

[44] C. He, G. Liu, H. Zhao, K. Zhao, Z. Ma, and X. An, Inorganic photovoltaic cells based on BiFeO$_3$: spontaneous polarization, lattice matching, light polarization and their relationship with photovoltaic performance, Phys. Chem. Chem. Phys. **22**, 8658 (2020).

[45] H. Yi, T. Choi, S. Choi, Y. S. Oh, and S. W. Cheong, Mechanism of the switchable photovoltaic effect in ferroelectric BiFeO$_3$, Adv. Mater. **23**, 3403 (2011).

[46] Z. Dai, A. M. Schankler, L. Gao, L. Z. Tan, and A. M. Rappe, Phonon-Assisted Ballistic Current from First-Principles Calculations, Phys. Rev. Lett. **126**, 177403 (2021).

[47] L. Z. Tan, F. Zheng, S. M. Young, F. Wang, S. Liu, and A. M. Rappe, Shift current bulk photovoltaic effect in polar materials-hybrid and oxide perovskites and beyond, Npj Comput. Mater. **2**, 1 (2016).

[48] S. M. Young and A. M. Rappe, First principles calculation of the shift current photovoltaic effect in ferroelectrics, Phys. Rev. Lett. **109**, 116601 (2012).

[49] T. Higuchi, Y.-S. Liu, P. Yao, P.-A. Glans, J. Guo, C. Chang, Z. Wu, W. Sakamoto, N. Itoh, T. Shimura, Electronic structure of multiferroic BiFeO$_3$ by resonant soft x-ray emission spectroscopy, Phys. Rev. B **78**, 085106 (2008).

[50] P. Baettig, C. Ederer, and N. A. Spaldin, First principles study of the multiferroics BiFeO$_3$, Bi$_2$FeCrO$_6$, and BiCrO$_3$: Structure, polarization, and magnetic ordering temperature, Phys. Rev. B **72**, 214105 (2005).